\documentclass[aps,pra,twocolumn,superscriptaddress,showkeys,longbibliography]{revtex4-1}	
\usepackage{amssymb, amsmath, amsbsy,array, latexsym}
\usepackage{graphicx}
\usepackage{dcolumn}
\usepackage{bm}
\usepackage{color}
\usepackage{mathrsfs}
\usepackage[colorlinks,urlcolor=blue,citecolor=blue,linkcolor=blue]{hyperref}
\usepackage{ragged2e}
\usepackage{multirow}
\usepackage[dvipsnames]{xcolor}

\usepackage{orcidlink}
\DeclareMathOperator{\sech}{sech}
\DeclareMathOperator{\csch}{csch}

\begin{document}

\preprint{APS/123-QED}

\title{Heat transport and rectification via quantum statistical and coherence asymmetries}


\author{Stephania Palafox\,\orcidlink{0000-0002-7585-9309}}
\author{Ricardo Román-Ancheyta\,\orcidlink{0000-0001-6718-8587}}
 \email{ancheyta6@gmail.com}
\affiliation{
 Instituto Nacional de Astrofísica, Óptica y Electrónica, Calle Luis Enrique Erro No.1 Santa María Tonantzintla, Puebla CP 72840, Mexico }
\author{Bar{\i}\c{s} \c{C}akmak\,\orcidlink{0000-0002-6124-3925}} 
\affiliation{College of Engineering and Natural Sciences, Bah\c{c}e\c{s}ehir University, Be\c{s}ikta\c{s}, Istanbul 34353, T\"urkiye}
\affiliation{TUBITAK Research Institute for Fundamental Sciences, 41470 Gebze, T\"urkiye}
\author{\"Ozg\"ur E. M\"ustecapl{\i}o\u{g}lu\,\orcidlink{0000-0002-9134-3951
}}
\affiliation{Department of Physics, Ko\c{c} University, \.{I}stanbul, 
Sar{\i}yer, 34450, T\"urkiye}
\affiliation{TUBITAK Research Institute for Fundamental Sciences, 41470 Gebze, T\"urkiye}

\date{\today}

\begin{abstract}

Recent experiments at the nanoscales confirm that thermal rectifiers, the thermal equivalent of electrical diodes, can operate in the quantum regime. We present a thorough investigation of the effect of different particle exchange statistics, coherence, and collective interactions on the quantum heat transport of rectifiers with two-terminal junctions. Using a collision model approach to describe the open system dynamics, we obtain a general expression of the nonlinear heat flow that fundamentally deviates from the Landauer formula whenever quantum statistical or coherence asymmetries are present in the bath particles. Building on this, we show that heat rectification is possible even with symmetric medium-bath couplings if the two baths differ in quantum statistics or coherence. Furthermore, the associated thermal conductance vanishes exponentially at low temperatures as in the Coulomb-blockade effect. However, at high temperatures it acquires a power-law behavior depending on the quantum statistics. Our results can be significant for heat management in hybrid open quantum systems or solid-state thermal circuits.

\end{abstract}

\maketitle


\section{Introduction}

The most wanted and ambitious goal in caloritronics~\cite{Giazotto_2017}, also known as thermotronics~\cite{Ben_2016}, is to have complete control~\cite{Casati_Chaos_2005} of the heat currents in micro and nanofabricated devices. Charge currents can interact with electromagnetic fields, and one can use this interaction to manipulate them. However, for heat currents carried phonons or photons~\cite{Pekola_Naure_2006}, the control of thermal transport is infeasible with standard techniques of modern electronics. Despite such difficulty, there has been impressive experimental progress in studying heat transport in atomic and molecular junctions~\cite{Longji_Cui_2017,Gotsmann_Nat_Nano_2017,Segal_Review_2016,Yonatan_RMP_2011}, nanostructures~\cite{Pop_Nano_Lett_2006}, and condensed matter systems~\cite{Pekola_RMP_2021} in the last two decades. Two representative examples are the observation of the quantum of thermal conductance~\cite{Schwab2000,Banerjee_2017} and the realization of solid-state heat valves~\cite{Pekola_Nat_Phys_2018,Winkelmann_PRL_2020} and thermal rectifiers (analogous to electrical diodes)~\cite{Chang_2006,Scheibner_2008,Giazotto_2015}. Thermal rectification is essential in electronics because it improves thermal management~\cite{ROBERTS2011648} and is closely related to thermal transistor effects~\cite{Miranda_Review_2017,Miranda_PRL_2016}.

Motivated by recent experiments with superconducting quantum circuits~\cite{Pekola,Pekola_Nat_Comm_2022} that validate long-standing theoretical models of quantum heat transport~\cite{Segal_PRL_2005,Segal_PRL_2009}, in this work, we contribute with a fundamental and insightful understanding of thermal rectification in such kinds of models. In particular, we thoroughly investigate the impact of quantum symmetry, quantum coherence, and collective interactions on the nonlinear heat transport of a central quantum system in contact with two baths/reservoirs/terminals.

We address the corresponding open system dynamics using a collision model approach~\cite{Ciccarello_Review_2022,Campbell}, also known as the repeated interactions scheme, of quantum thermodynamics~\cite{Myers2022,DeffnerCampbellBook,CP_Vinjanampathy}. We work in the continuous time limit of the collision model and obtain a time evolution in the Lindblad form, and describe various central systems (bosonic or fermionic) interacting with thermal and non-thermal baths in simple terms. By deriving a general heat flow expression, we identify unreported physical conditions and parameter regimes in which the system works as a quantum thermal rectifier. We also show that weak coherences in the reservoirs~\cite{Weakly,Chiara_Coherence_2022} induce thermal rectification in cases that otherwise would be impossible due to the lack of parametric and quantum statistical asymmetries. Furthermore, at low temperatures, we observe that the thermal conductance of the system vanishes exponentially, but it resembles that of crystals and carbon nanotubes at high temperatures. Practically, our results can be significant for heat management in compact hybrid quantum systems.

The paper's organization is the following. In Sec.~\ref{main}, we introduce the details of our model and present our fundamental results. Next, we discuss the implications of our general results in heat transport and rectification specific physical settings by assuming different quantum statistics for the medium and the bath particles in the Sec.~\ref{settings}. Sec.~\ref{conductance} presents the linear thermal conductance for the cases in Sec.~\ref{settings}. We conclude in Sec.~\ref{conclusion}.

\section{\label{main}Model of heat flow and fundamental results\protect}

\begin{figure}[t]
\centering
\includegraphics[height=4cm, width=7.5cm]{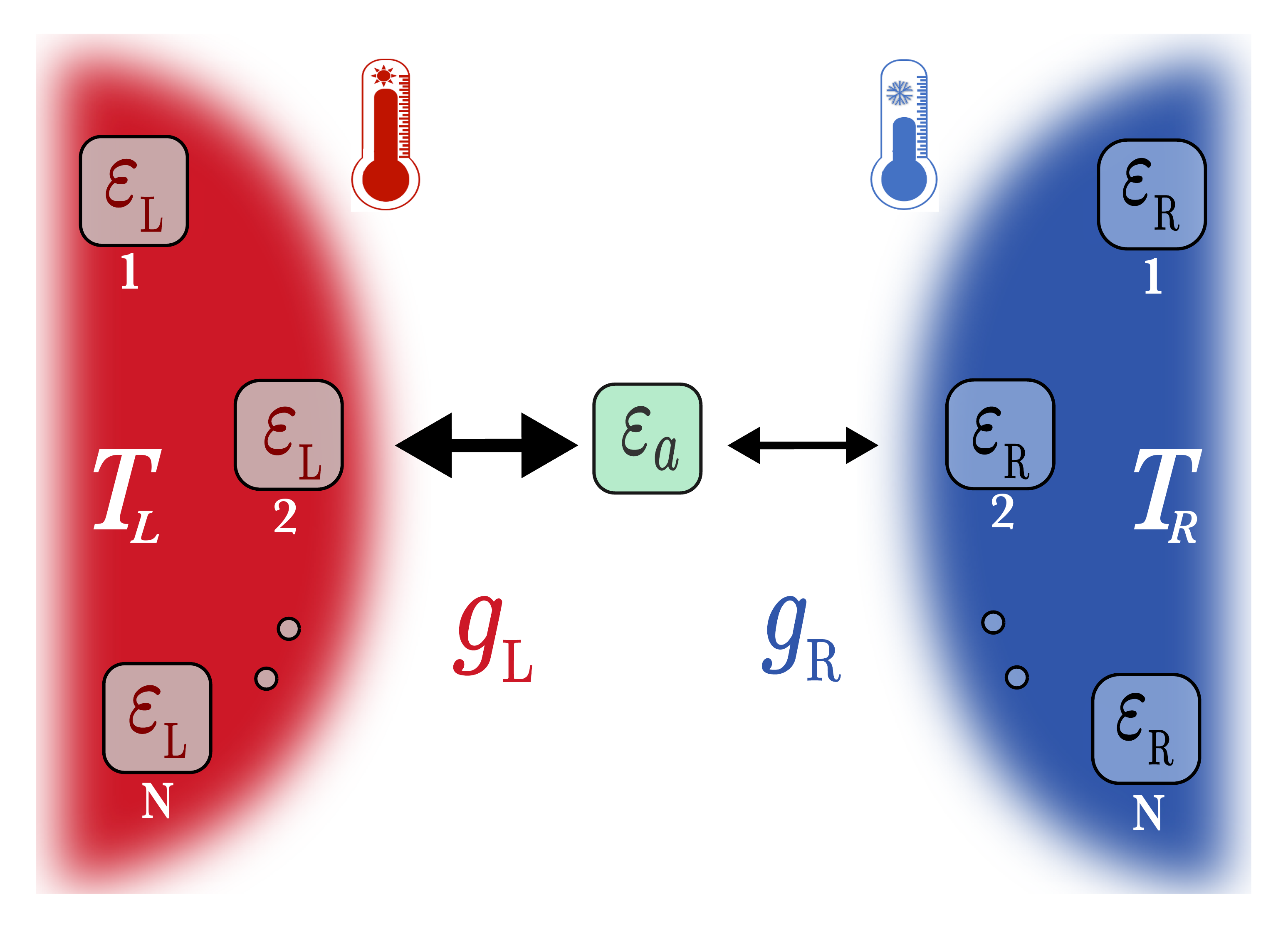}
\caption{Schematic representation of a quantum collision model~\cite{Ciccarello_Review_2022} describing heat transport through a thermal rectifier resembles an individual atom junction~\cite{Longji_Cui_2017,Gotsmann_Nat_Nano_2017}. The setup consists of a quantum system of interest (central green rectangle) asymmetrically coupled (black arrows) to two effective thermal baths at different temperatures ($T_L>T_R$). The labels $\varepsilon_a$ and $\varepsilon_{\lambda}^{}$ fix the quantum statistics of the central system and the baths [see Eqs.~(\ref{rules}) and (\ref{n_lambda})]. Even if the coupling coefficients $g_{\rm L}^{}$ and $g_{\rm R}^{}$ are equal, there will be thermal rectification if $\varepsilon_{\rm L}^{}\neq\varepsilon_{\rm R}^{}$ (see section \ref{hybrid_cases}) or if one of the baths contains a small amount of quantum coherence (see section \ref{Weakly_bath}).}
\label{general}
\end{figure}

We begin by introducing the physical model that we investigate in this work and present our central findings. Due to the diversity of the settings we are interested here, we adopt a versatile notation to express our results as compact as possible. To this end, we follow~\cite{CommutationRelations} and constrain the system and bath ladder operators to obey the following commutation or anticommutation relations
\begin{align}\label{rules}
    s s^\dagger-\varepsilon_a\, s^\dagger s &= 1 &\text{and} & & b^{}_{\lambda} b^\dagger_\lambda-\varepsilon_\lambda^{}\, b^\dagger_\lambda b^{}_{\lambda} &= 1,
\end{align}
where $s^\dagger$ ($s$) and $b^\dagger_\lambda$ ($b^{}_\lambda$) is the creation (annihilation) operators for the system and bath elements, respectively, and the subscript $\lambda={L}, {R}$ denotes the bath (left or right) label. While for $\varepsilon_a=1$ ($\varepsilon_\lambda^{}=1$) the system (bath) is of bosonic nature, in the case of $\varepsilon_a=-1$ ($\varepsilon_\lambda^{}=-1$) we have a fermionic system (bath). 
Throughout this work, particles of bosonic nature will be represented by quantum harmonic oscillators, and particles of fermionic nature will be represented by qubits (two-level quantum systems). Fig.~\ref{general} depicts a sketch of our physical model resembling a single-atom junction geometry~\cite{Longji_Cui_2017,Gotsmann_Nat_Nano_2017}.

In the context of collision models, it is possible to write the total Hamiltonian describing the global dynamics of the central system coupled to two effective thermal baths in the following general form
\begin{equation}\label{Total_H}
    H_T=H_S+H_L+H_R+H_{SL}+{H}_{SR}.
\end{equation}
Here, $H_S=\frac{1}{2}\hbar\omega(s^\dagger s+\varepsilon_a ss^\dagger)$, $H_\lambda=\frac{1}{2}\hbar\omega_\lambda(b_\lambda^\dagger b^{}_\lambda+\varepsilon_\lambda^{} b^{}_\lambda b_\lambda^\dagger)$, and depending on the value of $\varepsilon_a$ and $\varepsilon_{\lambda}$ these will be symmetric or antisymmetric self-Hamiltonians~\cite{Ancheyta_2011} of the central system and the  baths, where their natural frequencies are $\omega$ and $\omega_\lambda$, respectively. For instance, if $\varepsilon_a=-1$, $s^\dagger(s)$ should be replaced by the spin operator $\sigma_+(\sigma_-)$, this reduces $H_S$ to $\frac{1}{2}\hbar\omega\sigma_z$; the case $\varepsilon_a=1$ is easier to elucidate.  The last two terms in Eq.~(\ref{Total_H}) describe the interaction between the central system and the baths. For several physical scenarios, \textcolor{black}{for instance, to describe a realistic ultrasensitive calorimeter~\cite{Pekola_PRX_2022},} it is convenient to choose $H_{S\lambda}$ as the bilinear combination
\begin{equation}\label{bilinear_int}
{H}_{S\lambda}=g_\lambda^{}(s\,b_\lambda^\dagger +s^\dagger b^{}_\lambda),
\end{equation}
where $g_\lambda^{}$ is the coupling strength to the bath $\lambda$. \textcolor{black}{Strict energy conservation criteria is met whenever $[H_{S\lambda}, H_S+H_\lambda] =0$~\cite{Barra2015,Pereira,De_Chiara_2018}, and we have $\sum_\lambda[H_{S\lambda}\,,H_S+H_\lambda]=\sum_\lambda g_\lambda^{}(\omega-\omega_\lambda)(s b_\lambda^\dagger -s^\dagger b_\lambda^{})$ for any combination of $\varepsilon_a$ and $\varepsilon_\lambda^{}$.}
\textcolor{black}{Therefore, to simplify things further and to highlight the role that quantum coherence will have on the energetics of the system in Sec.~\ref{Weakly_bath}}, we assume that the central system and the baths are resonant with each other, i.e. $\omega=\omega_\lambda$. The resonant condition implies that the work cost of maintaining the collisions is strictly zero~\cite{Barra2015,Pereira,De_Chiara_2018,guarnieri2020non}.  Therefore, any energy change in the system is owing to energy flowing to or from the bath elements.
Following the well established steps of deriving the master equation for the central system from a collision model, 
we arrive at the Lindblad equation describing the corresponding open dynamics~\cite{Ciccarello_Review_2022,guarnieri2020non,PRA_Ancheyta_2021,roman2019spectral,Angsar19,Deniz19}
\begin{equation}\label{eq_mast}
     \dot{{\rho}}=-\frac{i}{2}\left[H_S,\rho\right]+\sum_\lambda g_\lambda^2\big(\langle b_\lambda b_\lambda^\dagger \rangle \mathcal{L}[s]\rho+\langle b_\lambda^\dagger b_\lambda\rangle  \mathcal{L}[s^\dagger]\rho\big),
\end{equation}
with $\mathcal{L}[x]\rho\equiv x\rho x^\dagger-\frac{1}{2}(x^\dagger x\rho+\rho x^\dagger x)$ being the usual Lindbland super-operator. Since we assume each bath element is in the thermal state $\rho_\lambda^{\rm th}=\exp\big(H_\lambda/k_B^{}T_\lambda\big)Z_\lambda^{-1}$ with $Z_\lambda$ being the partition function, then it is possible to express the expectation values appearing before the Lindblad super-operators ($\langle \mathcal{O}_\lambda\rangle\equiv{\rm tr}\{\rho_\lambda^{\rm th} \mathcal{O}_\lambda\}$) as follows $\langle b_\lambda^\dagger b_\lambda\rangle=n_\lambda^{}$ and $\langle b_\lambda b_\lambda^\dagger \rangle=1+\varepsilon_\lambda^{} n_\lambda^{}$,
where 
\begin{equation}\label{n_lambda}
n_\lambda^{}=\frac{1}{\exp\big(\hbar \omega/k_B^{} T_\lambda\big)-\varepsilon^{}_\lambda}.  
\end{equation}
If $\varepsilon_\lambda^{}=1$, $n_\lambda^{}$ is equal to the Bose-Einstein distribution, while for $\varepsilon_\lambda^{}=-1$ it is equal to the Fermi-Dirac distribution. $T_\lambda$ is the apparent temperature~\cite{Latune_2019} of bath $\lambda$ and $k_B^{}$ the Boltzmann's constant. In the high-temperature regime ($k_B^{}T_\lambda\gg\hbar\omega$) it is a simple exercise to show that $n_\lambda^{}$ can be well approximated by
\begin{equation}
n_\lambda^{}\approx \frac{\varepsilon_\lambda^{}}{2}\left[\left(\frac{2k_B^{}T_\lambda}{\hbar\omega}\right)^{\varepsilon_\lambda^{}}-1\right].
\end{equation}

The net heat current, $\mathcal{J}$, into (or out of) our central system is naturally composed of contributions coming from the left and the right bath, $\mathcal J=\mathcal J_{L}+\mathcal J_{R}$. It can simply be calculated using the standard definition~\cite{Kosloff_2013} $\mathcal{J}={\rm tr}\left\{H_S\mathcal{D}_{\rm diss}(\rho)\right\}$, where $\mathcal{D}_{\rm diss}(\rho)$ is the total dissipator given by the second term of Eq.~(\ref{eq_mast}) and yields the following result in full generality
\begin{equation}\label{J}
    \mathcal{J}={\sum}_\lambda\omega g_\lambda^2\left(n_\lambda^{}-\big[1+(\varepsilon_\lambda^{}-\varepsilon_a)n_\lambda^{}\big]\langle s^\dagger s\rangle\right).
\end{equation}
We are particularly interested in the steady-state behavior of this setup, which is attained when $\dot\rho=0$. The net or total heat current at the steady-state, $\mathcal{J}^{\rm SS}$ vanishes, i.e. $\mathcal{J}_{L}^{\rm SS}=-\mathcal{J}_{R}^{\rm SS}$, and $\langle s^\dagger s \rangle_{\rm SS} =(g_{\rm L}^2n_{\rm L}^{}+g_{\rm R}^2n_{\rm R}^{})/\sum_\lambda g^2_\lambda[1+(\varepsilon_\lambda^{}-\varepsilon_a)n_\lambda^{}]$.
As a result, it is possible to find a compact expression of the heat current from the left bath to the central system in the following way
\begin{equation}\label{J1}
\small
    \mathcal{J}^{\rm SS}_{L}=\omega g^2_{_{\rm L}} g^2_{_{\rm R}} \left \{ \frac{(n_{_{\rm L}}-n_{_{\rm R}})+(\varepsilon_{_{\rm R}}-\varepsilon_{_{\rm L}})n_{_{\rm L}} n_{_{\rm R}}}{g^2_{_{\rm L}}+g^2_{_{\rm R}}+g^2_{_{\rm L}} (\varepsilon_{_{\rm L}}\!-\varepsilon_a)n_{_{\rm L}}+g^2_{_{\rm R}} (\varepsilon_{_{\rm R}}\!-\varepsilon_a)n_{_{\rm R}}}\right\}.
\end{equation}
The equation above constitutes our first main result and expresses the nonlinear heat flow in terms of the statistics of the central system and bath elements, the system-bath couplings, and the temperatures of the baths through $n_\lambda^{}$. 

It is important to note that when we interchange the temperatures $T_L$ and $T_R$ in Eq.~(\ref{J1}), the only two terms capable of inducing an asymmetry in the heat flow are
$g^2_{\rm L} (\varepsilon_{\rm L}^{}\!-\varepsilon_a)n_{\rm L}^{}+g^2_{\rm R} (\varepsilon_{\rm R}^{}\!-\varepsilon_a)n_{\rm R}^{}$
in the denominator and $n_{\rm L}^{} n_{\rm R}^{}$ in the numerator. Furthermore, these terms are a direct consequence of the commutation relations [cf. Eq.~(\ref{rules})]; without them $\mathcal{J}_L^{\rm SS}(T_L,T_R)$ would be equal to $-\mathcal{J}_L^{\rm SS}(T_R,T_L)$. In the following sections, we will analyze the implications of such terms in various different system and bath combinations, and investigate the underlying mechanisms behind the rectification behavior in the heat flow.

Before moving on, we would like to point at an intriguing connection. It is possible to recast Eq.~(\ref{J1}) into a modified Landauer-type form describing heat transport across a two-terminal junction~\cite{Segal_PRL_2005}
\begin{equation}\label{Landauer}
\begin{split}
     \mathcal{J}_{L}^{\rm SS}=\int &\mathcal{T}(\nu)\left[n_{\rm L}^{}(\nu)-n_{\rm R}^{}(\nu)\right]\nu d\nu \\
     &+(\varepsilon_{\rm R}^{}-\varepsilon_{\rm L}^{})\!\!\int \mathcal{T}(\nu)\,n_{\rm L}^{}(\nu)n_{\rm R}^{}(\nu)\nu d\nu,     
\end{split}
\end{equation}
with the corresponding transmission coefficient being
\begin{equation}\label{transm_coeff}
\mathcal{T}(\nu)=\frac{g^2_{\rm L}\; g^2_{\rm R}\;\delta (\nu-\omega)}{g^2_{\rm L}\!+g^2_{\rm R}\!+g^2_{\rm L} (\varepsilon_{\rm L}^{}\!-\!\varepsilon_a)n_{\rm L}^{}(\nu)\!+g^2_{\rm R} (\varepsilon_{\rm R}^{}\!-\!\varepsilon_a)n_{\rm R}^{}(\nu)}.
\end{equation}
The first term in Eq.~(\ref{Landauer}) looks like the well-known expression of the Landauer formula, which indicates that heat conduction can also be viewed as a transmission phenomenon~\cite{Landauer_RMP_99}. 
The second (new) term in Eq.~(\ref{Landauer}) reflects the exchange statistics of the parts making up the physical setup. At a first glance, it may seem that any statistical asymmetry ($\varepsilon_{ \rm R}^{}\neq\varepsilon_{\rm L}^{}$) between the terminals, regardless of the temperature bias, could be sufficient to generate a nonzero energy flow; however, a close inspection shows that $\mathcal{J}_L^{\rm SS}=0$ whenever $T_L=T_R$. 
Unlike the Landauer–B\"uttiker formalism of quantum transport, here $\mathcal{T}(\nu)$ depends explicitly on the temperatures of the terminals through the function $n_{\lambda}^{}(\omega)$. This implies that $\mathcal{T}(\nu)$ contains, as we will see in the following sections, most of the rectification information of the system. 

\section{\label{settings}Different physical settings\protect}
\subsection*{Notation: ($\varepsilon_{\rm L}^{},\, \varepsilon_a,\, \varepsilon_{\rm R}^{}$)}

As outlined before, the physical settings we consider consists of a left bath, a central system and a right bath. They are embodied by either qubits or quantum harmonic oscillators (many of them in case of baths). In what follows, we will refer to these physical settings using the notation, ($\varepsilon_{\rm L}^{}, {\varepsilon_a}, \varepsilon_{\rm R}^{}$) (see Fig.~\ref{general}). 
For example, a setting where both baths and central systems are made out of qubits ($\varepsilon_\lambda^{}=\varepsilon_a=-1$) or harmonic oscillators ($\varepsilon_\lambda^{}=\varepsilon_a=1$) will be referred to as ($-1,-1, -1$) or ($1,1, 1$), respectively.

\subsection{Homogeneous cases: (-1, -1, -1) and (1, 1, 1)}\label{homo_sec}

We begin our discussion with arguably simplest settings in which both baths and the central system are embodied by either qubits or harmonic oscillators, which are denoted by ($-1,-1, -1$) or ($1,1, 1$) using the notation introduced before. As a figure of merit to assess possible thermal rectification performance of such settings we first adopt the so-called rectification coefficient defined by~\cite{landi2021nonequilibrium} $R\equiv\left \vert {J^{\rightarrow}}/{J^{\leftarrow}} \right \vert$.
Here $J^{\rightarrow}$ denotes a heat current induced by a temperature bias, $\Delta T=T_{L}-T_{R}$ hereafter $T_{L}>T_{R}$, that flows from the left bath towards the right bath. Then, $J^{\leftarrow}$ is the heat current when the temperatures of the thermal baths are interchanged, i.e. when the left (right) bath has a temperature equal to $T_{R}$ ($T_{L}$). Following these definitions, one can conclude that in case $R=1$, the system displays an entirely symmetric heat flow without any rectification at all, and when $R \rightarrow \infty$, the central system exhibits perfect rectification.

For the all qubit or all harmonic oscillator cases considered in this subsection, i.e. ($-1, -1, -1$) and ($1, 1, 1$), with the help of Eq.~(\ref{J1}) one can immediately conclude that $J^{\rightarrow}=-J^{\leftarrow}$ and therefore $R=1$, signaling the absence of thermal rectification. The reason behind this is both the second term in Eq.~(\ref{Landauer}) and $g^2_{\rm L} (\varepsilon_{\rm L}^{}\!-\varepsilon_a)n_{\rm L}^{}+g^2_{\rm R} (\varepsilon_{\rm R}^{}\!-\varepsilon_a)n_{\rm R}^{}$ vanish, making $\mathcal{T}(\nu)=g_{\rm L}^2\,g_{\rm R}^2\delta(\nu-\omega)/(g_{\rm L}^2+g_{\rm R}^2)$ temperature independent. As a result, the symmetry encompassing the cases where terminals and central systems have the same quantum statistics is sufficient to prevent any asymmetry in the heat flow~\cite{Segal_PRL_2009}. However, such assertion fails if at least one of the terminals possesses a small amount of quantum coherence, as we show in Sec.~\ref{Weakly_bath}.

\subsection{Baths with the same statistics: (1, -1, 1) and (-1, 1, -1) cases}\label{(1, -1, 1)}

We now turn our attention to a slightly different configuration in which we assume that the left and right thermal baths are made out of the same type of particles ($\varepsilon_{\rm L}^{}=\varepsilon_{\rm R}^{}$), but the central system has different quantum statistics than the baths ($\varepsilon_a\neq\varepsilon_{\lambda}^{}$). Two cases that fit this description are qubit baths with a harmonic oscillator as the central system ($-1,1, -1$), and harmonic oscillator baths with a central qubit system ($1,-1, 1$). 
Plugging the corresponding $\varepsilon_{x}$ values for the latter case in Eq.~(\ref{J1}) and evaluating the rectification coefficient one obtains~\cite{Pekola}
\begin{equation}
    R_{(1,-1, 1)}^{}=\frac{g^2_{\rm L} \coth{(\beta_{R}\, \omega/2)}+g^2_{\rm R} \coth{(\beta_{L}\, \omega/2)}}{g^2_{\rm L} \coth{(\beta_{L}\,  \omega/2)}+g^2_{\rm R} \coth{(\beta_{R}\, \omega/2)}}, \label{R_BQB}
\end{equation}
where $\beta_\lambda=(k_B^{}T_\lambda)^{-1}$ is the inverse temperature. For the former case the rectification coefficient is in the same form of Eq.~(\ref{R_BQB}) by replacing the $\coth (x)$ with $\tanh (x)$, which gives $R_{(-1,1,-1)}^{}=[R_{(1,-1, 1)}^{}]^{-1}$. 

A less common, but a more evident indicator and quantifier of the rectification behavior is called the {\it contrast} and follows~\cite{Pereira_EPL_2019,LandiContrast,Bibek_PRB_2021}
\begin{equation}\label{Contrast_Definition}
    C\equiv\left\vert\frac{ J^\rightarrow + J^\leftarrow}{J^\rightarrow - J^\leftarrow}\right\vert. 
\end{equation}
The contrast is bounded between zero and one, while $C=0$ indicates the absence of heat rectification, $C=1$ signals that the system is a perfect thermal diode. Such a definition resembles the Michelson contrast used in optics where $J^\rightarrow$ and $J^\leftarrow$ play the highest and lowest luminance role~\cite{Michelson_95}.
\begin{figure}[t]
\centering
\includegraphics[scale=0.4]{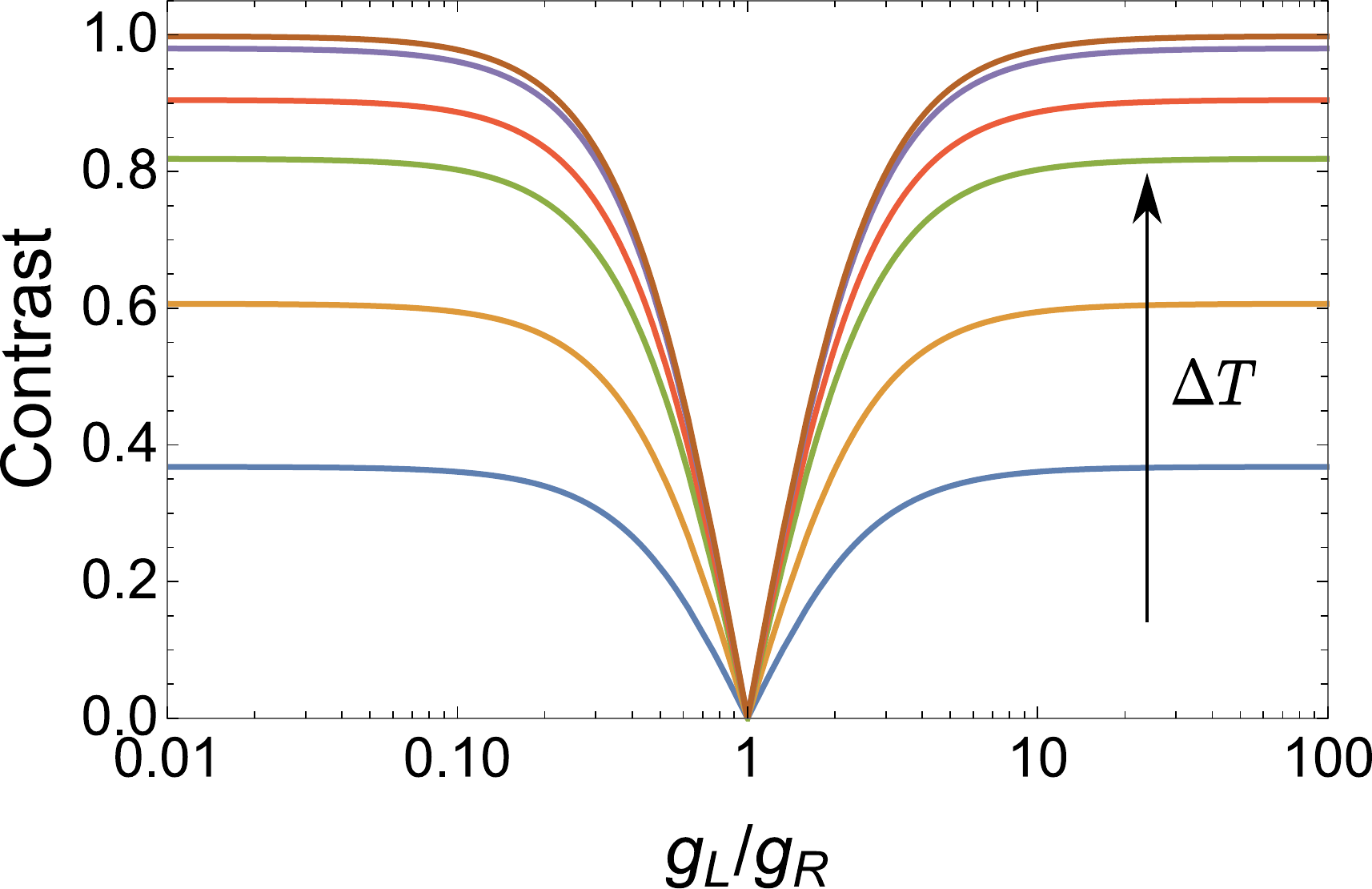}
\caption{The contrast [cf. Eq.~(\ref{C(-1,1,-1)})] quantifies thermal rectification as a function of the ratio of the bath couplings for the case where the baths have the same quantum statistics but different from the central system, i.e., $\varepsilon_{\rm L}^{}=\varepsilon_{\rm R}^{}\neq \varepsilon_a$. There is no rectification  ($C=0$) when the couplings to the baths are equal ($g_{\rm L}^{}=g_{\rm R}^{}$). Scaled bath temperatures are set as $k_B^{}T_{L}/\hbar\omega=1, 2, 5, 10, 50, 500$, and $k_B^{}T_{R}/\hbar\omega=0$ (see section \ref{(1, -1, 1)} for more details).}
\label{fig:sym_contrast}
\end{figure}
For the present physical setting, $C$ is calculated to give
\begin{equation}
    C=\left \vert\frac{g^2_{\rm L}-g^2_{\rm R}}{g^2_{\rm L}+g^2_{\rm R}}\right\vert \left\vert \frac{\coth{(\beta_{L}\,\omega/2)}-\coth{(\beta_{R}\,\omega/2)}}{\coth{(\beta_{L}\,\omega/2)}+\coth{(\beta_{R}\,\omega/2)}} \right \vert,
    \label{C(-1,1,-1)}
\end{equation}
where $C_{(1,-1, 1)}^{}=C_{(-1,1, -1)}^{}=C$.
Note that unlike Eq.~(\ref{R_BQB}), $C$ in Eq.~(\ref{C(-1,1,-1)}) consists of the product of two seemingly independent terms. While the first one only depends on the coupling parameters, the second one contains the information on the temperatures of the baths. It is easy to see that if $g_{\rm L}^{}=g_{\rm R}^{}$ or $T_L=T_R$ the system do not rectify the heat flow.

Figure~\ref{fig:sym_contrast} displays the behavior of the $C$ as a function of the coupling ratio $g_{\rm L}^{}/g_{\rm R}^{}$ for different values of the temperature difference, as given in Eq.~(\ref{C(-1,1,-1)}). Note that in the semi-log scale, $C$ is symmetric with respect to $g_{\rm L}^{}/g_{\rm R}^{}=1$, which is the only point that its value drops to zero provided that $\Delta T\neq 0$. This implies that it is possible to get heat rectification {only if} there is an asymmetry between the bath couplings, a condition known as ``parametric asymmetry'' first described in~\cite{Segal_PRL_2005} only for the $(1,-1,1)$ case.

We observe an advantage in considering baths and the central system with a different statistical nature than the previous all-qubits or all-harmonic oscillator settings where the rectification is absent. As measured by the contrast, the degree of rectification improves substantially as $\Delta T$ increases. Thus, it is instructive to look at the limit of large $\Delta T$ across the central system. For instance, in the high-temperature (classical) regime $\beta_\lambda^{}\hbar\omega/2\ll 1$, $\coth(\beta_\lambda^{}\hbar \omega/2)\approx 2k_B^{}T_\lambda^{}/\hbar\omega$, and $C\propto \Delta T$, i.e., thermal rectification becomes a linear function of the temperature bias. Conversely, in the low-temperature (quantum) regime $\beta_\lambda^{}\hbar\omega/2\gg 1$ and $\coth(\beta_\lambda^{}\hbar \omega/2)\approx 1$. Therefore, when $T_{L}\rightarrow\infty$ and $T_{R}\rightarrow 0$, we obtain the  following simple expression for the contrast
\begin{equation}
    C=\left \vert \frac{y^2-1}{y^2+1} \right \vert,\quad\text{with}\quad y=\frac{g_{\rm L}^{}}{g_{\rm R}^{}}. \label{high_deltaT}
\end{equation}
This result can also be obtained from Eq.~(\ref{C(-1,1,-1)}) simply by setting the second term that contains the temperatures equal to one. Also, Eq.~(\ref{high_deltaT}) is the simplified version of the asymmetry coefficient defined in~\cite{Bibek_PRB_2021} and measures spatial (parametric) asymmetry~\cite{Segal_PRE_2009}. The top line in Fig.~\ref{fig:sym_contrast} represents the behavior of the above equation, which shows that one could obtain a perfect thermal rectifier ($C=1$) in the current setting by introducing sufficient parametric asymmetry. There are also other works that present physical settings in which perfect rectification is achievable, however all of them have their central system made out of two or more interacting qubits~\cite{ContrastDefinition,Miranda_PRE_2017,Cahit_PRE_2019}, instead of the single central particle setting that we have in the present work. 

In relation to the experimental relevance of the presented results, we would like to highlight~\cite{Pekola}, where the authors recently displayed an experimental realization of the $(1,-1,1)$ case using a superconducting circuit, known as the spin-boson thermal rectifier~\cite{Segal_PRL_2005}. Here, in the context of thermal rectification and transport (see Sec.~\ref{conductance}), we find the thermally-equivalent, yet a slightly different scenario, the ($-1,1,-1$) case, that we call the boson-spin thermal rectifier in analogy with~\cite{Segal_PRL_2005}.


\subsection{Baths with different statistics: (1, -1, -1) and (1, 1, -1) cases}\label{hybrid_cases}

\begin{figure}[t]
{\bf (a)} \hskip4cm {\bf (b)}\\
\includegraphics[width=0.5\columnwidth]{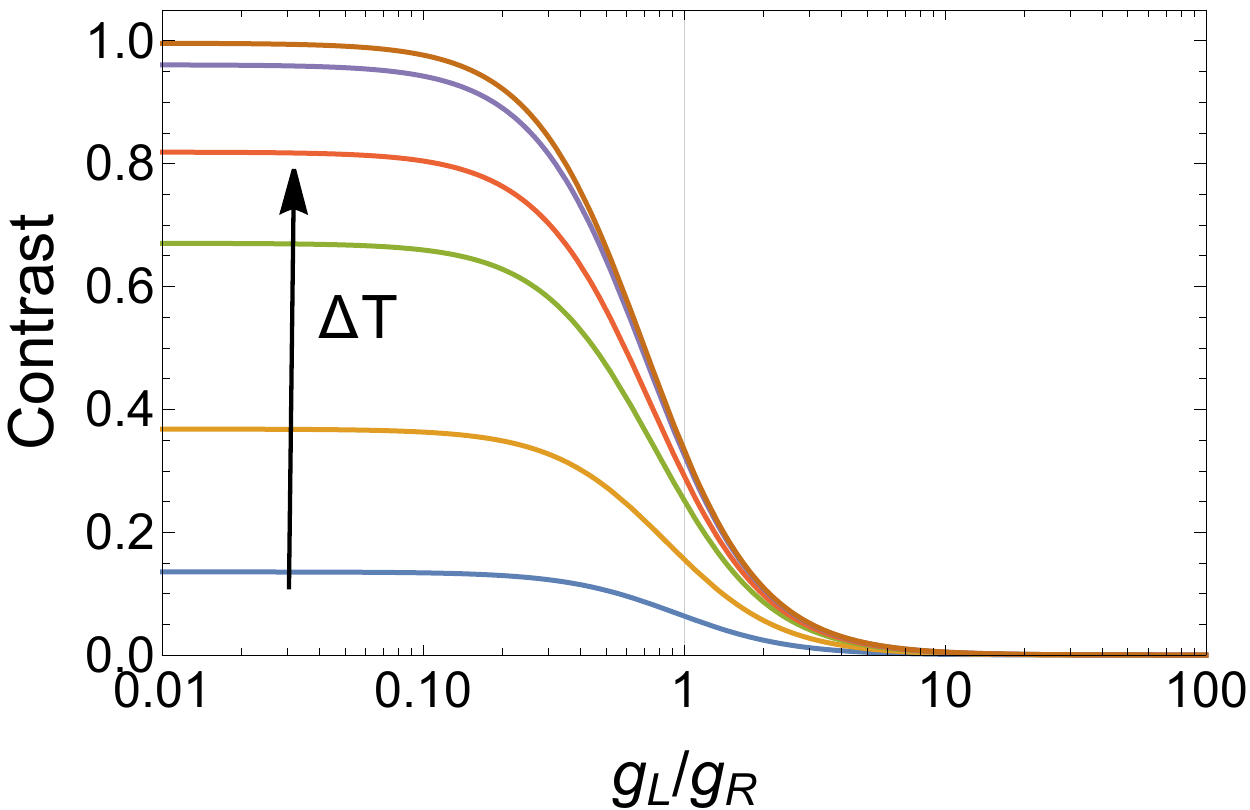}\includegraphics[width=0.5\columnwidth]{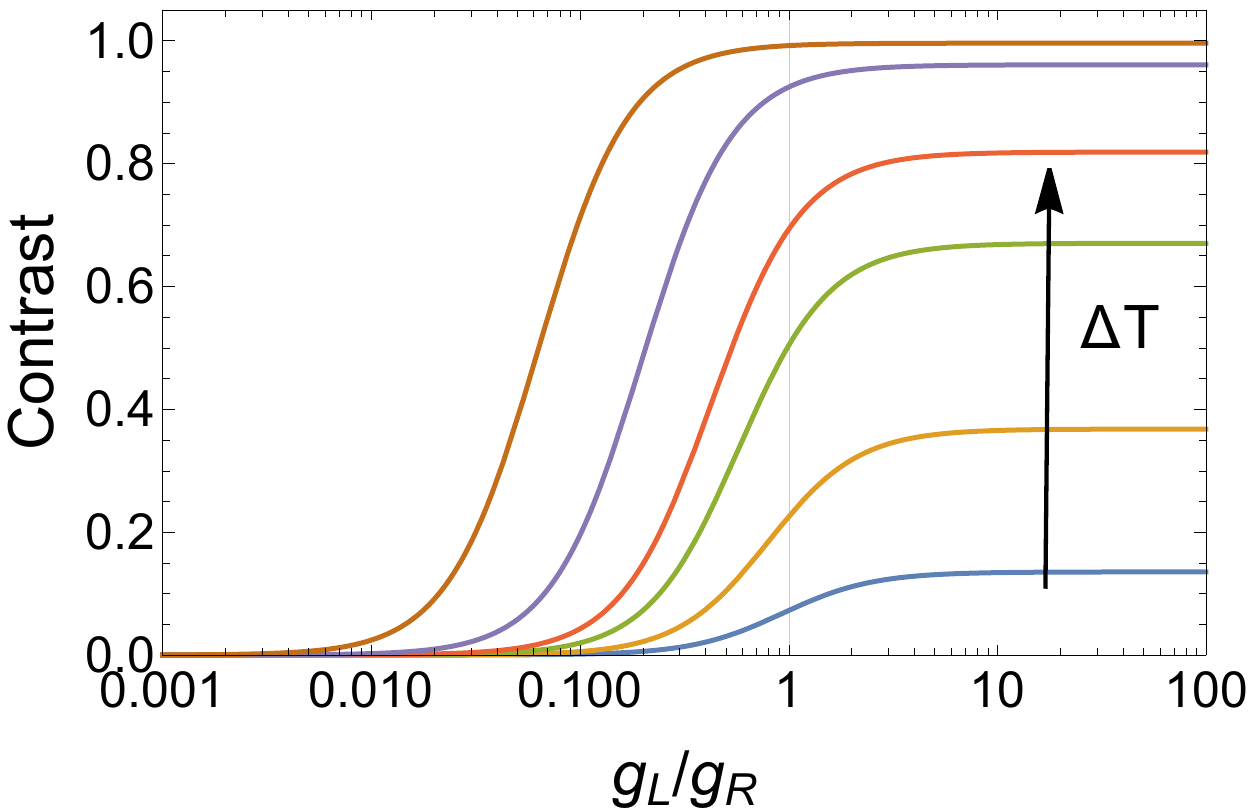}\\
\caption{Thermal rectification as a function of the ratio of the system-bath couplings for $(a)$ ($1,-1,-1$) and $(b)$ ($1,1,-1$) cases (see section \ref{settings} for notation definition) as quantified by the contrast given in Eqs.~(\ref{C(1,-1,-1)}) and (\ref{C(1,1,-1)}), respectively. In both panels the baths have different quantum statistics, i.e., $\varepsilon_{\rm L}^{}\neq\varepsilon_{\rm R}^{}$. Contrary to the case presented in Fig.~\ref{fig:sym_contrast}, we observe nonzero rectification at $g_{\rm L}^{}=g_{\rm R}^{}$ (faint vertical line); the effect is more pronounced for a harmonic oscillator central system (right panel). Scaled bath temperatures are set as  $k_B^{}T_{L}/\hbar\omega=1, 2, 5, 10, 50, 500$, and $k_B^{}T_{R}/\hbar\omega=0$ (see section \ref{hybrid_cases} for more details).}
\label{fig:asym_contrast}
\end{figure}

We continue with another variation of the two-terminal junction suitable to investigate hybrid quantum structures. We fix the left and right baths to be composed of harmonic oscillators and qubits, respectively, then analyze the cases of the central system being a qubit ($1,-1, -1$) or a harmonic oscillator ($1,1,-1$).  
We utilize the contrast as defined in Eq.~(\ref{Contrast_Definition}) to quantify the degree of rectification in these settings. The explicit expressions corresponding to the contrast are given as follows
\begin{equation}
    C_{(1,-1,-1)}^{}=\frac{\left|\tanh(\beta_{L}\,\omega/2)-\tanh(\beta_{R}\,\omega/2)\right|}{\tanh(\beta_{L}\,\omega/2)+\tanh(\beta_{R}\,\omega/2)+2g_{\rm L}^2/g_{\rm R}^2},
    \label{C(1,-1,-1)}
\end{equation}
%
\begin{equation}
    C_{(1,1, -1)}^{}=\frac{\left|\coth(\beta_{L}\,\omega/2)-\coth(\beta_{R}\,\omega/2)\right|}{\coth(\beta_{L}\,\omega/2)+\coth(\beta_{R}\,\omega/2)+2g_{\rm R}^2/g_{\rm L}^2}.
    \label{C(1,1,-1)}
\end{equation}
Left and right panels in Fig.~\ref{fig:asym_contrast} display the behavior of Eqs.~(\ref{C(1,-1,-1)}) and (\ref{C(1,1,-1)}), respectively. We observe thermal rectification depending on the ratio of bath coupling constants, which is already at variance with all qubits or harmonic oscillators cases. Therefore, it is more reasonable to compare the current setting with the previous one, Sec.~\ref{(1, -1, 1)}. The immediate and striking difference between the two is the nonzero value of the contrast at the point $g_{\rm L}^{}=g_{\rm R}^{}$ for both qubit and harmonic oscillator central particles. The mechanism behind the rectification in the present situation is the asymmetry generated by the dissimilar quantum statistics of the baths. The lack of parametric asymmetry in the case of equal coupling to the baths, which was necessary for rectification in the preceding subsection, is now compensated by such statistical asymmetry.  
Similar findings were obtained in~\cite{Segal_PRL_2009,Segal_PRE_2009} using a different and more elaborate procedure, where the core of their claim relies on identifying reservoirs with different mean energy~\cite{Segal_PRL_2009} or density of states~\cite{Segal_PRE_2009}. In contrast, our simple derivation of Eq.~(\ref{J1}) allows us to trace back the root of the statistical asymmetry to the commutation relations of Eq.~(\ref{rules}).
This result points out a potential use of particle statistics as a resource in quantum thermal machines, a topic of recent interest in the literature~\cite{Nathan1, Nathan2}.

When it comes to the differences between qubits and harmonic oscillators as central elements, there are two points that deserve special attention. First, note that in case of the qubit (harmonic oscillator) central system the parameter region that yields significant nonzero values for the contrast is $g_{\rm L}^{}\ll g_{\rm R}^{}$ ($g_{\rm L}^{}\gg g_{\rm R}^{}$). Keeping in mind that the left bath is composed of harmonic oscillators and right bath is composed of qubits, it is possible to conclude that finite contrast region for both cases is observed when the subsystem is more weakly coupled to the bath with the opposite statistical nature as compared to its coupling to the bath with the same statistical nature. Second, at the point (and below) of symmetric coupling, the oscillator subsystem performs significantly better than its qubit counterpart in rectifying the heat current [cf. the faint vertical line in Fig.~\ref{fig:asym_contrast} $(a)$ and $(b)$]. 
Thus, for a rectifier that is bound to operate between terminals composed of particles with different exchange statistics, it is convenient to design it with an oscillator central system to obtain better rectification when a symmetric coupling to the terminals is maintained. We believe this is a non-trivial observation, since often non-linearities or anharmonicities in the central element are thought to be advantageous in the classical~\cite{Pereira_EPL_2019} and quantum~\cite{Segal_PRB_2006} realms when it comes to achieving better performance in rectification. Keeping in mind that oscillators and qubits are linear and nonlinear systems, respectively, we have shown a counter-example of such a view.

Similar to the analysis in the previous subsection, it is instructive to look at the limit of extreme temperature bias with $T_{L}\rightarrow\infty$ and $T_{R}\rightarrow 0$, which yields
$C_{(1,-1, -1)}^{}=({1+2y^2})^{-1}$, with $y={g_{\rm L}^{}}/{g_{\rm R}^{}}$.
Once again we observe that the value of the contrast is only dependent on $g_{\rm L}^{}/g_{\rm R}^{}$ and the top most curve in Fig.~\ref{fig:asym_contrast} $(a)$ corresponds to this expression.
Considering the same large temperature gradient for the case ${(1,1, -1)}$ given in Eq.~(\ref{C(1,1,-1)}), we observe that it tends directly to one. Therefore, one may also obtain a perfect diode behavior with the $(1,1,-1)$ arrangement.

\textcolor{black}{On the experimental side, hybrid quantum circuits~\cite{RevModPhys.85.623} may be a promising platform for implementing the hybrid bath structures considered this subsection. For instance, the case $(1,-1,1)$ of previous Sec.~\ref{(1, -1, 1)} was implemented in~\cite{Pekola} by strongly coupling a central transmon-type qubit with two unequal superconducting co-planar waveguide resonators, each terminated with a copper microstrip resistor. We think one way to realize, e.g., the $(1,-1,-1)$ configuration, is by replacing one of the copper resistors mentioned above with an ensemble of electronic spins in the form of nitrogen-vacancy (NV) centers of a diamond crystal glued on top of the superconducting chip. Such spin ensembles can also be strongly coupled to a superconducting resonator~\cite{duty2010towards} or a flux qubit~\cite{Zhu2011} (see also~\cite{GREZES2016693}). In order to reveal how the statistical asymmetries between the baths act on the thermal rectification in such a platform, one should design the central system of interest, either a qubit or resonator, equally coupled to the hybrid baths, see Fig.~\ref{fig:asym_contrast}. Most likely, the main challenge would be manipulating the spin bath temperature with high precision. However, we hope our results will inspire experimentalists to focus on hybrid systems that control heat conduction by photons within the field of circuit quantum thermodynamics~\cite{Pekola2015NatPhy}.}



Furthermore, by following recent ideas of Refs.~\cite{roman2019spectral,PRA_Ancheyta_2021,Angsar19,Deniz19}, it is relatively easy to generalize previous results for the case where the central system interacts collectively with {\it clusters} of independent bath elements forming the terminals. In particular, when the left (right) bath is made of clusters of $N_{L}$ ($N_{R}$) qubits or harmonic oscillators, with $N_\lambda$ being an integer, one can show that the structure of the heat current as given in Eq.~(\ref{J1}) remains the same, but only the coupling coefficient $g_\lambda^{}$ is be replaced by the induced collective coupling given by $g_\lambda^{} \sqrt{N_\lambda}$~\cite{PRA_Ancheyta_2021,Deniz19}. When the clusters are of the same size, i.e. $N_L=N_R=N$, the heat current will be $N$ times $\mathcal{J}_L^{\rm SS}$ and the $C$ will be the same as before. However, in case one has $N_L\neq N_R$, the contrast can show a different behavior. For instance the high temperature gradient limit of the baths with the same statistics case given in Eq.~(\ref{high_deltaT}) is modified as $C=|(y^{*2}-1)/(y^{*2}+1)|$, where $y^{*2}\equiv g_{\rm L}^{2} N_L/g_{\rm R}^{2}N_R$. On the other hand, in the case of the baths with different statistics we have $C_{(1,-1,-1)}^{}=(1+2y^{*2})^{-1}$ in the large $\Delta T$ limit. Therefore, even when the parametric asymmetry is absent ($g_{\rm L}^{}=g_{\rm R}^{}$), collective interactions can still induce thermal rectification.

\subsection{(-1, -1, -1) case with a weakly coherent bath} \label{Weakly_bath}

As the last physical setup, we re-visit the configuration in which both baths and the central system of interest are composed of qubits ($\varepsilon_\lambda=\varepsilon_a=-1$) but with a slight twist. Without loss of generality, we assume that the density matrices of the elements composing the left bath are no longer genuinely thermal; instead, they have small but nonzero off-diagonal elements, i.e. weak coherences. We have already shown that in the case of all qubits configuration with thermal baths, heat flow is entirely symmetric presenting no rectification behavior (see Sec.~\ref{homo_sec}). Our aim here is to see if thermal rectification can be observed due to the presence of weak coherences in one of the baths. If true, the superposition of quantum states can be considered a new form of asymmetry sufficient to obtain rectification.
\begin{figure}[t]
\includegraphics[scale=0.4]{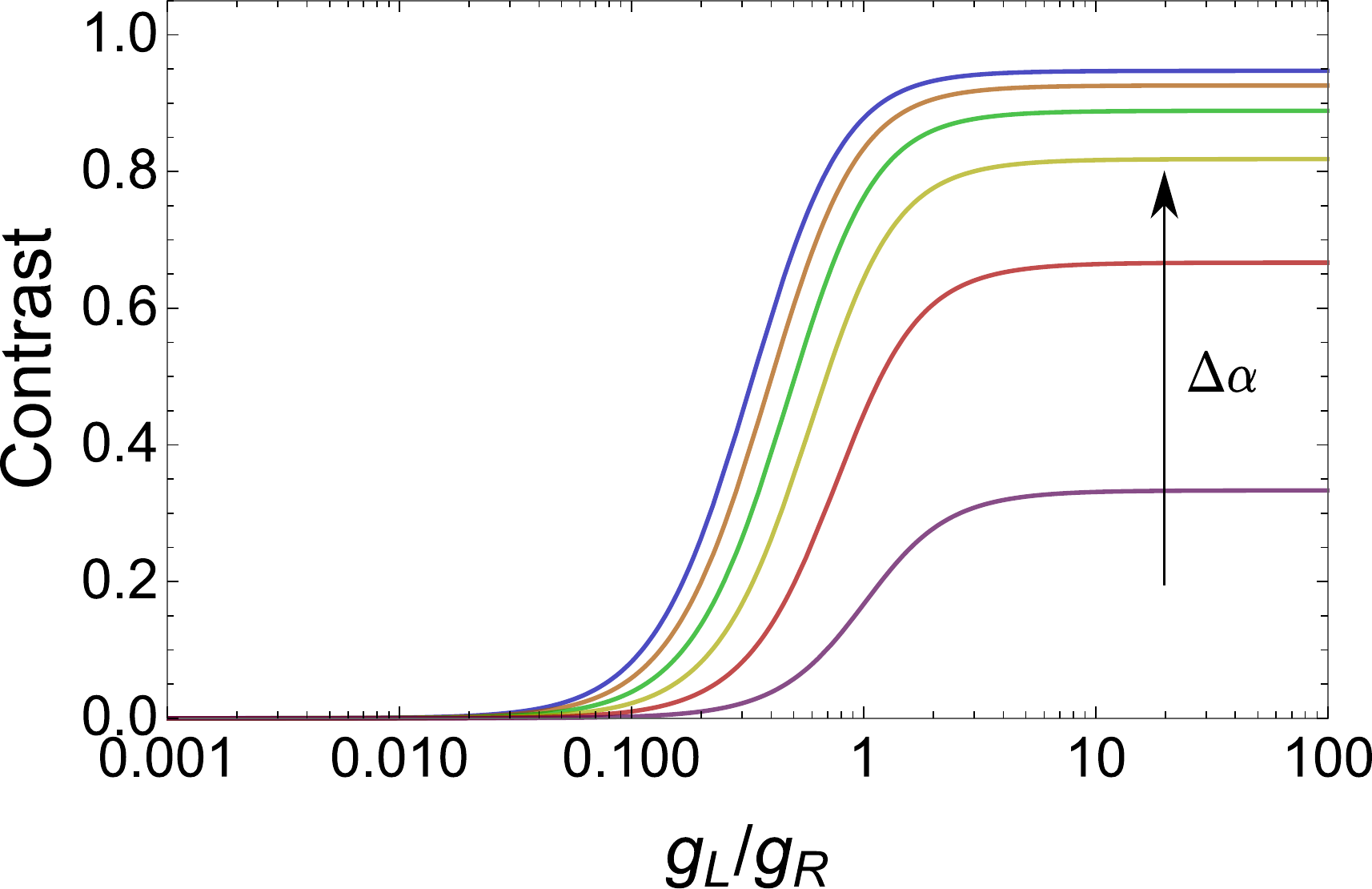}
\caption{Contrast for ($-1,-1,-1$) case with weakly coherent left bath and thermal right bath as a function of the ratio of the bath couplings. $\alpha$ increases from the bottom line to the top one, we set $\alpha=0.25, 0.5, 0.75, 1, 1.25, 1.5$. We observe that the effect of increasing a coherence bias is similar to that of increasing the temperature bias in case of both baths are thermal.}
\label{fig:weaklycoherent_contrast}
\end{figure}

Following closely the framework of {\it weakly coherent} collisional models, recently introduced in~\cite{Weakly}, we begin by writing the initial state of the bath elements of the left terminal in the following form
\begin{equation}
    \rho_{L}^{}=\rho_{L}^{\rm th}+\sqrt{\tau}\alpha\chi,  \label{Rho_Coherence}
\end{equation}
where $\chi$ is a Hermitian operator that represents the coherent part of the density matrix with $\alpha$ characterizing the magnitude of the coherences, and $\tau$ is the interaction time between the central system of interest and the bath elements. Note that, the limit of weak coherences is attained when the second term in the above equation is much smaller than the first one. This can be achieved in the continuous time limit of the model when $\tau\rightarrow 0$, which is a necessary step in the derivation of the master equation governing the dynamics of the central system given below (see \cite{Weakly} for details). The bath elements in the right bath are in a thermal state, $\rho_{R}^{\rm th}$, as generally done throughout this work. 
The master equation describing the dynamics of the central qubit in the present situation is given in the following form~\cite{Weakly,Chiara_Coherence_2022}
\begin{equation}
    \dot{\rho}=-\frac{i}{2}\big[H_S+\alpha\,
    G\,,\,\rho\big]+\mathcal{D}_{\rm diss}(\rho),\label{Master_Equation_Coherence}
\end{equation}
where $\mathcal{D}_{\rm diss}(\rho)$ is the total dissipator given by the second term of Eq.~(\ref{eq_mast})
that is associated with the thermal parts $\rho_\lambda^{\rm th}$, and $G={\rm tr}_{L}^{}\{H_{SL}^{}\,\chi\}$ represents an additional unitary contribution to the subsystem dynamics originating from the presence of coherences in the left bath. For the sake of simplicity and to highlight the role of quantum coherence we choose the coherent part to be in the form $\chi=|0\rangle \langle 1|+|1\rangle \langle 0|$, so $G=g_{\rm L}^{}\sigma_x$.
To calculate the change in the internal energy we resort to the usual definition $d\langle H_S\rangle/dt={\rm tr}\{ \dot{\rho}H_S\}$, keeping in mind that $H_S$ is independent of time.
Using the master equation (\ref{Master_Equation_Coherence}) the (modified) first law now splits into three contributions~\cite{Weakly}
\begin{equation}
\begin{split}
    d\langle H_S\rangle/dt=\dot{\mathcal{W}}_{C}+\mathcal J_{L}+\mathcal J_{R}, 
    \label{first_law_modify}
\end{split}
\end{equation}
where $\dot{\mathcal{W}}_C\equiv i\alpha\langle[G,H_S]\rangle=\frac{1}{2}\omega\alpha g_{\rm L}^{}\langle \sigma_y\rangle$ is the coherent work and $\mathcal J_{\lambda}=\omega g^2_\lambda(n_\lambda^{}-\langle \sigma_+\sigma_- \rangle)$ the incoherent heat. 
%
At the steady-state the left-hand side of Eq.~(\ref{first_law_modify}) vanishes as expected; however, due to the coherent work, $\mathcal{J}_L^{\rm SS}$ will no longer be equal to $-\mathcal{J}_R^{\rm SS}$. Thus, unlike all previous cases, the contrast (and also the rectification coefficient) will depend on whether we calculate it using the left or right heat current, which in practice depends on the ability to measure one or the other. In particular, the explicit form of $\mathcal{J}_L^{\rm SS}$ is
\begin{equation}
\begin{split}
    \mathcal J_{L}^{\rm SS}=\frac{\omega g_{\rm L}^2 g_{\rm R}^2}{g_{\rm L}^2+g_{\rm R}^2}(n_{\rm L}^{}-n_{\rm R}^{})
    -
    \frac{\omega g_{\rm L}^{2}\alpha^{*2}}{g_{\rm L}^2+g_{\rm R}^2}\sum_\lambda g_\lambda^2 \tanh(\beta_\lambda^{}\omega/2),
    \label{Q_Incoherent}
\end{split}
\end{equation}
where
$\alpha^*\equiv{\alpha g^{}_{\rm L}}{\left[\omega^2+(g^2_{\rm L}+g^2_{\rm R})^2+2\alpha^2 g^2_{\rm L}\right]^{-1/2}}$.
%
The second term of Eq.~(\ref{Q_Incoherent}) is a clear modification in the heat current due to quantum coherence ($\alpha\neq 0$) in the elements forming the left, non-thermal bath. When $\alpha=0$, we would have recovered the form that lead us to a symmetrical heat current, as presented in Sec.~\ref{homo_sec}.

At this point, we are ready to calculate the rectification performance of the current setting as quantified by the contrast~(\ref{Contrast_Definition}). In order to make a more concise analysis of the situation, we are going to reduce the number of free parameters describing our system by fixing some of them. From the previous examples, we have established that in a given physical setting, the higher the temperature difference between the baths, the higher the contrast is. Therefore, we concentrate on this large temperature bias limit, $\Delta T\rightarrow\infty$, and try to shed some light on the effect of coherence bias between the baths. In this case, the steady-state contrast is given by
\begin{equation}
    C=\left[{1+\frac{g^2_{\rm R}}{\alpha^2 g^2_{\rm L}}\left( \frac{\omega^2}{g^2_{\rm L}+g^2_{\rm R}}+g^2_{\rm L}+g^2_{\rm R} \right)}\right]^{-1}.
    \label{Contrast_Q_Incoherent}
\end{equation}

Figure \ref{fig:weaklycoherent_contrast} displays the behavior of the contrast expression above, for different values of of the left bath coherences. Notice how the coherence difference between the baths, $\Delta \alpha=\alpha_L^{}-\alpha_R^{}=\alpha_L^{}$, plays the role of the temperature difference played in the previous examples of fully thermal baths. More importantly, even when the system-bath couplings are equal to each other, i.e. $g^{}_{\rm L}=g^{}_{\rm R}$, the asymmetry provided by the coherences in the left bath particles is sufficient to induce thermal rectification. This further proves that the asymmetry necessary for the rectification can be provided by means other than the coupling of the central system to the baths. On the other hand, it also presents a clear advantage provided by quantum coherence when used as an additional resource for designing thermal management devices~\cite{Naseem_PRR_2020,PRE_Li,Kasper_Alan_2021} and complements recent efforts showing that asymmetry in the internal coherence of a two qubits central system is sufficient for the emergence of thermal rectification~\cite{Naseem_PRE_2021}. 

\textcolor{black}{It is important to note that the thermal rectification discussion in this section is in context with the analysis made in the previous sections since our calculations only involve the incoherent heat flowing through the central system, which unambiguously satisfies the laws of thermodynamics for weakly coherent baths~\cite{Weakly}}


\section{Thermal Conductance}\label{conductance}

Another convenient and complementary way to analyze the heat transport is through the thermal conductance, which, in the linear response regime, is defined as~\cite{Segal_JCP_2003,Yamamoto_NJP_2018}
$\kappa=\lim^{}_{\Delta T\rightarrow 0}\ {\mathcal{J}}/\Delta T.$
Following from our general result for $\mathcal{J}$ given in Eq.~(\ref{J1}), we obtain
\begin{equation}\label{Thermal_Conductance}
    \kappa=\frac{\omega^2 g_{\rm L}^2 g_{\rm R}^2 n_{\rm L}^{}n_{\rm R}^{} \exp\left({\hbar\omega/k_{B}^{}T}\right)}{k_B^{} T^2\left[g^2_{\rm L}+g^2_{\rm R}+g^2_{\rm L}(\varepsilon_{\rm L}^{}-\varepsilon_a)n_{\rm L}^{}+g^2_{\rm R}(\varepsilon_{\rm R}^{}-\varepsilon_a)n_{\rm R}^{}\right]}. 
\end{equation}
Using the above expression, it is possible to obtain $\kappa$ for the specific cases considered in the previous sections. For instance, when both terminals have the same quantum statistics but they are different from that of the central system, i.e., $\varepsilon_{\rm L}^{}=\varepsilon_{\rm R}^{}\neq\varepsilon_a$, the conductance is
\begin{equation}\label{Conduct_symetric_cases}
\kappa=\frac{g^2_{\rm L} g^2_{\rm  R}\omega^2}{g^2_{\rm L}+g^2_{\rm R}}\frac{\csch\left(\hbar\omega/k_B^{}T\right)}{2k_B^{}T^2},
\end{equation}
and as function of the temperature it exhibits a Schottky-type behavior~\cite{Yamamoto_Conds_Matr_2021}. It is important to note that, similar to what we found for the contrast in Sec.~\ref{(1, -1, 1)}, Eq.~(\ref{Conduct_symetric_cases}) is the same for both $(1, -1, 1)$ and $(1, -1, 1)$ configurations. This result reinforces the fact that the spin-boson thermal rectifier, which is widely studied in the literature~\cite{Segal_PRL_2005}, is somehow thermally equivalent to its boson-spin counterpart presented here. At low temperatures ($k_B^{}T\ll \hbar\omega$), Eq.~(\ref{Conduct_symetric_cases}) [and also Eq.~(\ref{Thermal_Conductance})] predicts an exponential suppression of the heat transport, which has been interpreted as the thermal analogous of the Coulomb-blockade effect occurring in electrical
transport~\cite{Saito_PRL_2013}. For instance, electrical conductance is exponentially suppressed in a quantum dot due to the increased charging energy and the Coulomb repulsion between individual electrons within the device~\cite{Beenakker_PRB_1991}. Here, the vanishing of $\kappa$ can be understood due to the lack of thermal excitations in the hot bath to populate the excited states of the central system which, after relaxation, should deposit its energy into the cold bath producing a sequential tunneling process from the left to the right terminal.~\cite{Yamamoto_NJP_2018}. 
On the other hand, at high temperatures ($k_B^{}T\gg \hbar\omega$) Eq.~(\ref{Conduct_symetric_cases}) reduces to
\begin{equation}
\kappa\approx\frac{\omega g^2_{\rm L} g^2_{\rm R}}{g^2_{\rm L}+g^2_{\rm R}} \frac{1}{2k_B^{}T}.
\end{equation}
Interestingly, the $1/T$ dependence coincides with the expected behavior of the thermal conductance in crystals, where at high temperatures, the Umklapp phonon-phonon  scattering~\cite{Ziman2001} is the dominant process that can cause thermal resistance~\cite{Yoshimura_PRE_2022}.

Table~\ref{table:Thermal_Conductance} displays Eq.~(\ref{Thermal_Conductance}) for the rest of the cases studied in Secs.~\ref{homo_sec} and \ref{hybrid_cases}. As we mentioned above, $\kappa$ has exponential suppression at low temperatures independently of the value of $\varepsilon_x$, but it exhibits distinct behaviors at high temperatures. For instance, when the baths and central system are all harmonic oscillators, $\mathcal{J}_L^{\rm SS}$ describes ballistic heat transport, and $\kappa$ is independent of the temperature~\cite{Segal_PRE_2009}.  However, when all are qubits (highly anharmonic systems), the thermal conductance behaves as $1/T^{2}$. Remarkably, this quadratic decay has also been observed in individual single-wall carbon nanotubes above room temperature~\cite{Pop_Nano_Lett_2006}. When the baths have different quantum statistics ($\varepsilon_{\rm L}^{}\neq\varepsilon_{\rm R}^{}$), one of the coupling coefficients ($g_\lambda^{}$) in $\kappa$ becomes temperature-dependent, see Table~\ref{table:Thermal_Conductance}. In addition, from the discussion of collective interactions at the end of Sec.~\ref{hybrid_cases}, we expect a linear increase in the thermal conductance with the clusters' size.

Finally, for perfectly adiabatic contact between reservoirs, the transmission coefficient $\mathcal{T}(\nu)$ [cf. Eq.~(\ref{transm_coeff})] reduces to one, and together with the first term of Eq.~(\ref{Landauer}), it is relatively  easy to show that the quantum of the thermal conductance is $\pi^2k_B^2 (T_{L}+T_{R})/6h$~\cite{PRL_Rego_98}, a value that has been experimentally demonstrated in suspended nanowires at very low temperatures~\cite{Schwab2000} or, more recently, in single-atom junctions at room temperature~\cite{Longji_Cui_2017,Gotsmann_Nat_Nano_2017}. 

\begin{table}[h]
\centering
\resizebox{8.5cm}{!}{
\begin{tabular}{ c c | c c c | c c }
\hline

\multicolumn{2}{|c|}{$(\,\varepsilon_{\rm L}^{},\, \varepsilon_a\,, \varepsilon_{\rm R}^{}\,)$} &
\multicolumn{3}{|c|}{Thermal conductance} & \multicolumn{2}{c|}{High temperature}\\
\hline \hline

\multicolumn{2}{|c|}{} & \multicolumn{3}{c|}{} & \multicolumn{2}{c|}{}\\

\multicolumn{2}{|c|}{$(+1,+1,+1)$} & \multicolumn{3}{c|}{\(\displaystyle\frac{\omega^2 g^2_{\rm L} g^2_{\rm R}}{g^2_{\rm L}+g^2_{\rm R}}\frac{\csch^2\!\big(\hbar\omega/
2k_B^{}T\big)}{4k_B^{2}T^2}\)} & \multicolumn{2}{c|}{\(\displaystyle\frac{g^2_{\rm L} g^2_{\rm R}}{g^2_{\rm L}+g^2_{\rm R}}\)}\\
\multicolumn{2}{|c|}{} & \multicolumn{3}{c|}{} & \multicolumn{2}{c|}{}\\

\multicolumn{2}{|c|}{$(+1,+1,-1)$} & \multicolumn{3}{c|}{\(\displaystyle\frac{\omega^2 g^2_{\rm L} [g^{}_{\rm R}(T)]^2}{g^2_{\rm L}+[g^{}_{\rm R}(T)]^2} \frac{\csch^2\!\big(\hbar\omega/
2k_B^{}T\big)}{4k_B^{2}T^2}\)} & \multicolumn{2}{c|}{\(\displaystyle \omega g^2_{\rm R} \frac{1}{2k_B^{} T}\)}\\

\multicolumn{2}{|c|}{} & \multicolumn{3}{c|}{} & \multicolumn{2}{c|}{}\\

\multicolumn{2}{|c|}{$(-1, -1, - 1)$} & \multicolumn{3}{c|}
{\(\displaystyle
\frac{\omega^2 g^2_{_{\rm L}} g^2_{_{\rm R}}}{g^2_{_{\rm L}}+g^2_{_{\rm R}}}\frac{\sech^2\!\big(\hbar\omega/
2k_B^{}T\big)}{4k_B^2T^2}
\)} & \multicolumn{2}{c|}{\(\displaystyle
\frac{\omega^2 g^2_{\rm L} g^2_{\rm R}}{g^2_{\rm L}+g^2_{\rm R}}\frac{1}{4k_B^2T^2}\)}\\
\multicolumn{2}{|c|}{} & \multicolumn{3}{c|}{} & \multicolumn{2}{c|}{}\\

\multicolumn{2}{|c|}{$(+1, -1, -1)$} & \multicolumn{3}{c|}{\(\displaystyle\frac{\omega^2 [{g}^{}_{\rm L}(T)]^2 g^2_{\rm R}}{[{g}^{}_{\rm L}(T)]^2+g^2_{\rm R}} \frac{\sech^2\!\big(\hbar\omega/
2k_B^{}T\big)}{4k_B^{2}T^2}\)} & \multicolumn{2}{c|}{\(\displaystyle \omega^2 g^2_{\rm R} \frac{1}{4k_B^{2} T^2}\)}\\

\multicolumn{2}{|c|}{} & \multicolumn{3}{c|}{} & \multicolumn{2}{c|}{}\\

\hline
\end{tabular}
}
\caption{Linear thermal conductance [see Eq.~(\ref{Thermal_Conductance})] and its corresponding high-temperature limit for distinct combinations of $\varepsilon_x$. For the cases where the baths have different statistics ($\varepsilon_{\rm L}^{}\neq\varepsilon_{\rm R}^{}$) we have defined the temperature dependent coupling coefficients ${g}^{}_{\rm R}(T)\equiv g^{}_{\rm R} \tanh^\frac{1}{2}(\hbar\omega/2k_B^{}T)$ and $g^{}_{\rm L}(T)\equiv g^{}_{\rm L}\coth^\frac{1}{2}(\hbar\omega/2k_B^{}T)$.
}
\label{table:Thermal_Conductance}
\end{table}

\label{K(-1,-1,-1)HT}


\section{Conclusions}\label{conclusion}

We have investigated the thermal rectification and conductance behavior of a quantum medium (harmonic oscillator or qubit) functioning between two baths, and we obtain useful analytical results. First, we systematically analyzed the role that the intrinsic exchange symmetry of the medium and the bath particles have on the thermal rectification performance. For instance, we found that the boson-spin thermal rectifier is thermally equivalent to its spin-boson counterpart, which is well studied in the literature (see Sec.~\ref{(1, -1, 1)}). For both rectifiers, an asymmetric coupling to the baths (parametric asymmetry) is not sufficient to observe perfect rectification; one also needs to consider a significant temperature bias to achieve it. On the contrary, parametric asymmetry is not required for hybrid cases in which the hot and cold bath particles have different quantum statistics. Interestingly, in such cases, a rectifier with a harmonic oscillator as the central element outperforms the one using a qubit system, which is highly anharmonic, for all parameter regimes. We also found that collective interactions can asymmetrically increase the coupling strength to the baths; therefore, they can be used to obtain thermal rectification when parametric asymmetry is not a viable option. Remarkably, for cases where the central system and the baths have identical quantum statistics, we have shown that quantum coherence in at least one bath can induce rectification, which is otherwise impossible regardless of the system parameters (see Sec.~\ref{Weakly_bath}).  

At a more fundamental level, we have developed a modified Landauer-type expression for the heat current in our physical model [see Eq.~(\ref{Landauer})], which explicitly includes the quantum statistics of the central system and the bath elements. Taking advantage of the general expression of the heat current, we have derived the linear thermal conductance for different combinations of the central system and bath particles' quantum statistics. We have found that while the conductance is the same for all possible combinations at the low-temperature limit, it exhibits different power-law behaviors for different combinations in the high-temperature limit.
We note that the latter results may be relevant for crystal structures and carbon nanotubes above room temperature, which are experimentally shown to display similar behaviors. 
Finally, one could generalize Eq.~(\ref{Landauer}) and subsequent  results to the case of q-deformed algebras~\cite{Arik_1976,sym1020240}, where the q-commutation relation is $a_qa^\dagger_q-q\, a_q^\dagger a_q=1$, but $q$ is now a real parameter in the region $0<q<1$. However, one quickly realizes that the expectation value $\langle a^\dagger_q a_q\rangle$ will substantially differ from Eq.~(\ref{n_lambda}), see~\cite{TUSZYNSKI_1993,Lavagno_PRE_2002}. Therefore, such a generalization is far from trivial and outside the scope of this work.

Our results on the quantum statistical and coherent effects in heat transport can fundamentally shed light on the interplay between quantum mechanical properties of the building blocks of a thermal machine and its energetics. Furthermore, the presented results can be experimentally relevant in heat transport in compact hybrid systems where the variation of temperature gradient over the small system sizes can be challenging.

\acknowledgments
S.P. would like to express her gratitude to CONACyT, Mexico for her Scholarship. B. \c{C}. is supported by The Scientific and Technological Research Council of Turkey (TUBITAK) under Grant No.~121F246 and BAGEP Award of the Science Academy.



%




\newpage

\end{document}